\documentclass{aa}  

\usepackage{graphicx}
\usepackage{txfonts}
\usepackage{textcomp}
\usepackage{soul}
\usepackage{epstopdf}
\usepackage{longtable}
\usepackage{hyperref}
\hypersetup{colorlinks,allcolors=black}
\usepackage{xcolor}

\begin{document}

   \title{Broadband radio study of the supernova remnant Kes~73}


   \author{S. Loru
          \inst{1}, 
          A. Ingallinera\inst{1}, 
          A. Pellizzoni\inst{2}, 
          E. Egron\inst{2}, 
          C. Bordiu\inst{1}, 
          G. Umana\inst{1}, 
          C. Trigilio\inst{1}, 
          F. Bufano\inst{1}, 
          M.N, Iacolina\inst{3},          
          M. Marongiu\inst{2},          
          S. Mulas\inst{2},
          C. Buemi\inst{1},
          F. Cavallaro\inst{1},
          P. Leto\inst{1},
          A. Melis\inst{2},
          P. Reich\inst{4},
          W. Reich \inst{4},
          S. Riggi\inst{1}
          and\\ 
          A.C. Ruggeri \inst{1}
          }

   \institute{$^{1}$ INAF, Osservatorio Astrofisico di Catania, Via Santa Sofia 78, 95123 Catania, IT\\
              \email{sara.loru@inaf.it}\\      
   $^{2}$ INAF, Osservatorio Astronomico di Cagliari, Via della Scienza 5, 09047 Selargius, Italy\\
   $^{3}$ ASI, Osservatorio Astronomico di Cagliari, Via della Scienza 5, 09047 Selargius, Italy\\
   $^{4}$ Max-Planck-Institut f\"ur Radioastronomie, Auf dem H\"ugel 69, 53121 Bonn, Germany          
         %
             }

   \date{}

  \abstract
{Strong shocks occurring in supernova remnants (SNRs), and their interaction with an often anisotropic surrounding medium, make SNRs ideal laboratories for studying the production and acceleration of cosmic rays (CRs). Due to their complex morphology and phenomenology, different CR populations are expected to exist throughout the remnants, each characterized by its own energy spectrum. A comprehensive understanding of particle acceleration mechanisms and energetics in SNRs requires spatially resolved spectral and morphological studies.
}
{We want to highlight the crucial role of high-resolution radio images at high frequencies ($\gtrsim 10$~GHz) for studying the spectral properties of different remnant regions and better constraining the models that describe their non-thermal emission from radio to $\gamma$-ray wavelengths.}
{
We studied the integrated radio spectrum of the SNR Kes~73 using single-dish observations performed with the Sardinia Radio Telescope (SRT) between $6.9$ and $24.8$~GHz, complemented by published data. The high-resolution map at $24.8$~GHz was used to search for spatial variations in the spectral index across the remnant. 
}
{We present the SRT images of Kes~73, providing the highest-frequency morphological and spectral characterization ever obtained for this source. By combining our 18.7 and 24.8 GHz maps with previously published interferometric images at 1.4 and 5~GHz, we identify a flatter spectrum in the western bright region compared to the rest of the shell.
In the same region, we detect overlapping $^{12}$CO molecular emission and $\gamma$-ray radiation, providing strong evidence of SNR–molecular cloud interaction and enhanced CR production. We modelled the non-thermal radio to $\gamma$-ray emission from this region, favouring a lepto-hadronic scenario with a maximum electron energy of $1.1$~TeV and a magnetic field strength of $25\,\mu$G.}
{The spatial coincidence of the radio-bright, flat-spectrum region with the $^{12}$CO emission, together with the preferred lepto-hadronic model, strongly supports an ongoing interaction between the Kes~73 shock front and nearby molecular clouds.}

   \keywords{ISM: supernova remnants  -- ISM: individual objects: Kes~73 --
                 Radio continuum: general --
                Radiation mechanisms: non-thermal 
               }
   
   \titlerunning{Radio study of the SNR Kes~73}
   \authorrunning{S. Loru et al.}
 \maketitle
%

\section{Introduction}
\label{Sec:Introduction}

Supernova remnants (SNRs) are highly complex objects in which expanding ejecta carry an enormous amount of mechanic energy ($\sim 3\times 10^{50}$~erg; \citealt{Leahy_2020}). This energy budget is transferred in $\sim 1$~kyr to the magnetic field, and to the kinetic and thermal energies of the shocked interstellar gas and particles, which are accelerated up to relativistic speeds (\citealt{Truelove_1999}, \citealt{Leahy_2019}). The cosmic ray (CR) electrons and protons thus produced are responsible for the non-thermal continuum radiation observed from radio to $\gamma$-ray frequencies. 

In the radio band, synchrotron emission from relativistic electrons is the dominant radiation mechanism, 
typically resulting in a power-law spectrum (flux density $S_{\nu} \propto \nu^{\alpha}$, with $\alpha \sim -0.5$). 
Deviations from this trend can occur at low frequencies (below $\sim 100$~MHz), where thermal absorption can introduce a turnover, and at high frequencies (above $\sim 10$~GHz),
where a spectral cutoff can be observed in connection with the maximum energy of the accelerated electrons.
The cutoff frequency, as well as the integrated spectral index, is strongly dependent on the SNR’s evolutionary stage
and provides critical constraints on particle acceleration models \citep{Urosevic_2014}.
Both hadronic ($\pi^0$-decay emission) and leptonic processes (inverse Compton and bremsstrahlung) contribute to the $\gamma$-ray spectrum. 
Since the leptonic components are produced by the same relativistic electrons responsible for the radio synchrotron emission, 
the radio spectral slope and high-frequency cutoff can constrain the leptonic contribution to the observed $\gamma$-ray emission.

Extended and complex SNRs often contain regions subject to different physical conditions, influenced by shock properties, variations in the surrounding interstellar medium (ISM) and circumstellar medium, and possible interactions with atomic or molecular clouds (MCs; \citealt{Egron_2017}, \citealt{Loru_2018}, \citealt{Loru_2024}). These factors affect the particle-acceleration mechanisms, leading to distinct populations of relativistic electrons across the remnant, each with its own energy distribution and characteristic non-thermal spectrum.  
In particular, the co-spatial presence of bright, flat-spectrum radio emission and $\gamma$-ray emission indicates regions of enhanced CR acceleration.  
In this context, accurate estimates of the integrated spectral index, including spatially resolved spectral analyses, along with X-ray and infrared (IR) emissions recovering the thermal dust emission \citep{SNR_book_2020}, are crucial for achieving a comprehensive understanding of SNRs and their radiation mechanisms \citep{Dubner_2015}.
Therefore, sensitive flux-density measurements over a wide frequency range (ideally $\sim 0.1$–$100$~GHz), combined with high-resolution imaging, are needed to resolve regions affected by different shock conditions and local environmental properties (\citealt{Egron_2017}, \citealt{Loru_2018}). 
This approach has been limited by technical challenges in obtaining both sensitive and high-resolution observations of extended sources at radio and $\gamma$-ray energies.
Consequently, most studies of SNR non-thermal emission have relied on radio flux densities integrated over the entire remnant, leading to misinterpretations of theoretical models. 
A significant advance in these studies can be achieved through single-dish observations with large-diameter telescopes, which combine sensitive flux density measurements of extended sources with high-resolution imaging up to high radio frequencies.

Kes~73 (SNR~G27.4+0.0) is one of the youngest Galactic SNRs, with an estimated age of  500 to 1000~yr (\citealt{Tian_2008}, \citealt{Borkowski_2017}).
It is classified as a shell-type remnant with a radius of $\sim 4^{\prime}$ \citep{Green_catalogue_2022}. Its distance was estimated to be between 7.5 and 9.8~kpc based on H~\textsc{i} absorption studies \citep{Tian_2008}. 
In a more recent study, \citet{Ranasinghe_2018} used H~\textsc{i} observations to revise the distance of Kes~73 to 5.8~kpc. This value has been confirmed by near-IR H$_2$ line emission observations  
\citep{Lee_2020}. 
At its centre, the remnant harbours the magnetar 1E~1841$-$045, well detected in X-rays \citep{Krumar_2014}, which is believed to be the compact remnant of the Kes~73 progenitor star (\citealt{Krumar_2014}, \citealt{Castelletti_2021}).

The radio emission from Kes~73 appears slightly asymmetric, with a brighter region along the western side of the shell \citep{Ingallinera_2014}. 
Radio studies have revealed a steep integrated spectrum ($\alpha = -0.68$; \citealt{Green_catalogue_2019}) and spatial spectral variations in the 1.4–5 GHz range, 
with significantly higher indices (up to $\alpha\sim-0.2$) towards the bright western region \citep{Ingallinera_2014}. 
A spectral turnover also was observed below $\sim 100$~MHz, likely due to absorption by ionized gas associated with a MC interacting with the remnant \citep{Castelletti_2021}. 

Kes~73 is also clearly detected in the mid-IR band, particularly at $24\,\mu$m,  where the emission morphology closely matches the radio structure. 
This mid-IR emission was attributed to heated dust by \citet{Priestley_2021}. 
\citet{Castelletti_2021} reported a spatial coincidence between the 74-MHz radio and $24$ $\mu$m emission from the western shell and a $^{13}$CO MC in the $95$–$105$~km\,s$^{-1}$ velocity range, supporting the hypothesis of an SNR–MC interaction.

XMM and \textit{Chandra} observations presented by \citet{Krumar_2014} revealed a strong morphological correspondence between the X-ray, radio, and IR emissions, particularly in the western limb, which is the brightest region at all wavelengths. 
An extended \textit{Fermi}-Large Area Telescope (LAT) source was proposed as the $\gamma$-ray counterpart of Kes~73 in the 0.1–300~GeV range (`source A' in \citealt{Liu_2017}), 
with its centroid located near the western bright region. 

This multi-wavelength scenario, from radio to X-ray domains, emphasizes the western limb as a region of interest, given its association with both molecular material and high-energy emission. 
These characteristics make it a compelling candidate for a CR-acceleration site.

In this work we present new single-dish observations of Kes~73 performed with the Sardinia Radio Telescope (SRT) between $6.9$ and $24.8$~GHz, which reach the highest radio frequency ever explored for this source. 
These images enabled, for the first time, a detailed morphological analysis of Kes~73 at high radio frequencies,
as well as an investigation of both its integrated and spatially resolved spectral behaviour in a previously unexplored frequency domain. 
We exploited the high resolution of our single-dish images, which extend up to 24.8~GHz, to accurately constrain the non-thermal spectrum of the western radio-bright region of Kes~73 across radio and $\gamma$-ray wavelengths. This enabled us to gain valuable insights into the energetics of the CRs accelerated in this region and to estimate the strength of the magnetic field.

In Sect.~\ref{sec:Observations and data analysis} we describe the SRT observations and the main steps of the data analysis. 
The ancillary  Murchison Widefield Array (MWA) and MeerKAT datasets used for the spectral and morphological characterization of Kes~73 are presented in Sect.~\ref{Ancillary data}. 
The results, including final calibrated images, flux-density measurements, and spectral-index maps, are given in Sect.~\ref{sec:Results}.
In Sect.~\ref{Sec:5} we model the Kes~73 non-thermal emission from the radio to $\gamma$ rays and present the corresponding results (Sect.~\ref{Sec: Multiwavelength study}); we also briefly discuss the complex region surrounding Kes~73 (Sect.~\ref{sec:ThecrowedKes73field}). Finally, we summarize our conclusions in Sect.~\ref{sec:Conclusions}.

\section{SRT observations and data analysis}
\label{sec:Observations and data analysis}

We observed Kes~73 with SRT between December 2018 and January 2019, at central frequencies of 6.9~GHz ($C$ band) and 18.7, 22.2, and 24.8~GHz ($K$ band). These observations were conducted as part of the SRT observing programme\footnote{Programme code SRT 22-18, PI: A. Pellizzoni.} dedicated to the study of spatially resolved spectral indices up to high radio frequencies for a sample of both young and middle-aged SNRs \citep{Loru_2021}.  Additional observations were carried out in June 2021 during the SRT Director’s Discretionary Time (DDT) programme\footnote{DDT programme SRT June 2021, PI: A. Pellizzoni.}, aimed at $K$-band monitoring of the SNRs Cas~A and Kes~73. The spectro-polarimetric backend SARDARA (SArdinia Roach2-based Digital Architecture for Radio Astronomy; \citealt{Melis_2018}) was used to record data in full-Stokes mode, with 1024 spectral channels and 1.4 GHz bandwidth for both the $C$ and $K$ bands. We created on-the-fly (OTF) maps with a size of $0.5^{\circ} \times 0.5^{\circ}$, significantly larger than the extent of Kes~73, to ensure a proper background baseline subtraction.
Observations of Kes~73 were always carried out at elevations above $20^{\circ}$, ensuring optimal SRT performance and effectively minimizing pointing errors and beam shape instabilities.
Both the $C$- and $K$-band observations were carried out with a scanning velocity of $4^{\prime}\mathrm{sec^{-1}}$ and a sampling interval of 20~ms. We set the offset between two consecutive scans to $0.6^{\prime}$ at 6.9~GHz and  $0.2^{\prime}$ at 18.7, 22.2, and 24.8~GHz, implying about four passages per beam size at all observing frequencies. 
In the $C$ band, these parameters imply about  $153$~sample beam$^{-1}$ and an exposure time of $3$~sec beam$^{-1}$.

We used the seven-feed $K$-band receiver in the best space coverage configuration (\citealt{Bolli_2015}) to perform the observations at 18.7, 22.2 and 24.8~GHz. 
As described in \citet{Loru_2018}, this geometry is well suited for mapping extended sources since it allows us to obtain: (i) a beam oversampling, necessary to accurately evaluate the continuum flux densities, and (ii) a large improvement in mapping speed that minimizes the impact of temporal atmospheric opacity variations on the map quality. Taking advantage of these characteristics, we obtained $K$-band maps with at least $\sim 244$~sample beam$^{-1}$ (referred to the highest frequency), by considering the contribution of all seven feeds, and a minimum exposure time of $4.9$~sec beam$^{-1}$ for a full RA or Dec map. The quality of the $K$-band maps presented here is ensured by the recommended opacity conditions ($\tau$<0.1 neper) that we verified during our `shared-risk mode' observations.  
The observations and related parameters are summarized in Table \ref{tab:obs_table}. 

We used the SRT Single-Dish Imager (SDI) software (\citealt{Egron_2016}, \citealt{Prandoni_2017}, \citealt{Loru_2018}, \citealt{Marongiu_2020_methods}) to perform the data reduction. SDI was developed to reduce single-dish data acquired with all SRT receivers and backends through 
an automatic pipeline (quicklook analysis) and interactive tools for data inspection, baseline removal, radio frequency interference (RFI) rejection and image
calibration. A detailed description of the $C$- and $K$-band SDI data reduction is given by \citet{Egron_2017} and \citet{Loru_2018}. The data acquired with the  SARDARA backend are reduced through both a `spatial' and a `spectral' RFI rejection procedure (as described in \citealt{Egron_2017}), after which the data are averaged into a single continuum channel on which the following data reduction steps are applied.
The final products of the SDI pipeline are Flexible Image Transport System (FITS) images suitable for scientific
analysis tools. 
From an accurate evaluation of  spectrum in terms of RFI and spectral features and their impact on the map quality, we rejected the spectral channels at the frequencies of $6.36-6.41$~GHz and $21.64-21.98$~GHz from the $C$-band and $K$-band (at the central frequency of $22.2$ GHz) data, respectively. 
We produced the images shown in this paper in units of Jy~beam$^{-1}$, and using the cubehelix colour scheme \citep{Green_2011}. 

To apply the data calibration, we carried out cross-scan observations on the point-like sources 3C286, 3C295, 3C147, 3C48, 3C123 and NGC~7027, and we reconstructed or interpolated their flux density at the observed frequencies by using flux densities and the polynomial expressions proposed by \cite{Perley_2013}, \cite{Perley_2017}, and \cite{Zijlstra_2008}. 
In order to guarantee the consistency between calibrators and Kes~73 in terms of the gain stability and observing condition, repeated cross-scan observations on calibrators were performed during each observing session. With the same aim, in the data reduction phase, the calibration factors were selected so that: (i) they have same backend attenuation
parameters as the target, and (ii) their observations are performed at least within 12 h (or less in the case of changing weather) from each target scan epoch and possibly at a similar elevation.

\begin{table*}
\caption{
SRT observations and related results in terms of integrated flux density and rms.}
\label{tab:obs_table}
\centering
\begin{tabular}{|c|c|c|c|c|c|c|c|c|}
\hline\hline
Epoch & N.maps & Obs. time & Freq. & HPBW & Flux density & rms & calib. err & Exc. feeds \\
 &  &  (h) & (GHz) & ($^{\prime}$) & (Jy) & (mJy/beam) & & \\
\hline
2018 Dec 19 & 7 & 1.59 & 6.9  & 2.75 & $1.65 \pm 0.05$ & 8  & 2.5\%  & - \\
            & 2 & 1.14 & 22.2 & 0.83 & $1.15 \pm 0.06$ & 10 & 4.8\%  & 3R, 5L, 6L \\
2019 Jan 25 & 7 & 4.0  & 24.8 & 0.75 & $0.86 \pm 0.09$ & 5  & 10.3\% & 3R, 5L, 6L \\
2021 Jun 14 & 4 & 2.30 & 18.7 & 0.99 & $0.88 \pm 0.04$ & 7  & 4.3\%  & 3R, 4L, 5L, 6R \\
2021 Jun 15 & 2 & 1.15 & 24.8 & 0.75 & $0.85 \pm 0.05$ & 12 & 5.7\%  & 3R, 4L, 5L, 6R \\
\hline
\end{tabular}
\tablefoot{With `N.maps', `Obs. time', `Freq.' and `HPBW',  we indicate the number of maps, the total observation time (including overheads), the central frequency and the half-power beam width. A single map is intended as a complete scan (RA or Dec) on the source.  We also report the calibration errors and the feed channels (indicated with the feed number and `L' and `R' for the left and right polarization channels)  discarded from data analysis (due to Low Noise Amplifiers problems).}
\end{table*}

Differently from the mono-feed
$C$-band receiver, the $K$-band receiver consists of seven feeds. For this reason, during each $K$-band observing session, we also created OTF maps of calibrators in order to take the single-feed specific efficiency into account (\citealt{Orfei_2010,Loru_2018}). We used these maps to calculate the ratios between the expected flux of the calibrator and the peak counts related to each feed, and we corrected  each single-feed map of Kes~73 by applying these scaling factors. 
In the case of $K$-band data, we first analysed and separately calibrated each single-feed/polarization channel map. Through both visual inspection and signal-to-noise evaluation of each map, we flagged the images related to the feed/polarization channels reported in Table \ref{tab:obs_table} for each observing session, and we discarded them from our analysis in order to optimize the final image quality. 

For both $C$-band and $K$-band images, 
the high number of bright and extended unrelated sources in the field prevented  
a precise baseline subtraction. This results in a background noise contribution that we subtracted by following the procedure described in Sect. \ref{sec:Results}.

We calculated the flux-density uncertainties by adding in quadrature the calibration errors and the statistical errors.  
The calibration errors account for both the standard deviation of the calibration factors derived for each observing session and the uncertainties associated with the beam model. These include slight variations in the half-power beam width due to changes in elevation and weather conditions, and are estimated to be $\sim5\%$ for the $K$ band, while they are negligible for the $C$-band data.
We calculated the statistical uncertainties as
$\sigma~\times (N_\mathrm{beams})^{0.5}$, where  $\sigma$ is the standard deviation
associated with the background region and $N_\mathrm{beams}$ is the number of beam solid angles contained in the extraction area of the target. We note that calibration errors are definitely dominant for our SRT $K$-band observations. Related percentage uncertainties are reported in Table \ref{tab:obs_table} for each frequency and observing session. 


\section{Ancillary data}
\label{Ancillary data}
\subsection{MWA data}
\label{subsec:MWA data}

Kes~73 was observed with MWA as part of the GaLactic and Extragalactic All-sky MWA (GLEAM; \citealt{Wayth_2015}) survey. 
We downloaded public MWA cutouts centred on Kes~73, available via the GLEAM Postage Stamp Service\footnote{\url{http://gleam-vo.icrar.org/gleam_postage/q/form}}, 
corresponding to four `wideband' images, each with a bandwidth of approximately 30~MHz and central frequencies of 88, 118, 155, and 200~MHz. The complementary frequency coverage of the GLEAM images with respect to the SRT data makes them particularly useful for accurately evaluating the integrated spectral index of Kes~73 and the turnover due to thermal absorption. 

The MWA data are affected by a background noise contribution, 
which we estimated and subtracted following the same procedure described for the SRT data in Sect.~\ref{sec:Results}. We computed the integrated flux density of Kes~73 using aperture photometry on each image. Following \citet{Hurley-Walker_2017}, 
the flux-density calibration uncertainties for the sky region around Kes~73 are $\sim 8$\%. 
We computed the total uncertainties on our flux-density measurements by adding this contribution and the statistical errors in quadrature, 
mainly associated with background subtraction (we followed the same procedure as for the SRT data in Sect.~\ref{sec:Observations and data analysis}).

\subsection{MeerKAT data}
\label{subsec:SMGPS data}

Kes~73 falls within the SARAO MeerKAT legacy Galactic
Plane Survey (SMGPS; \citealt{Goedhart_2024}), which consists of $L$-band continuum observations ($886$–$1678$~MHz) of a large portion of the Galactic Plane
($2^{\circ} < l < 60^{\circ}$, $252^{\circ} < l < 358^{\circ}$, and $|b| < 1.5^{\circ}$). 
From the public SMGPS first data release archive\footnote{\url{https://doi.org/10.48479/3wfd-e270}},
we retrieved the Kes~73 image at the central frequency of $1.284$~GHz. 
The image has a resolution of $8^{\prime\prime}$ and an rms noise level of $\sim 50~\mu$Jy/beam.

Kes~73 is included in the SMGPS extended source catalogue \citep{Bordiu_2025}, where a flux density of $4.57 \pm 0.23$~Jy is reported. This value is consistent with the flux densities at $0.960$ \citep{Trushkin_1996} and 1.4~GHz \citep{Ingallinera_2014} previously reported in the literature (see Table \ref{tab: flux densities}).
Nevertheless, we noted that prominent artefacts introduced during the imaging process significantly affect the tile containing Kes~73. 
Bright sources appear aliased, with spurious features displaced by a few arcminutes from their true positions. 
While these issues only have a minor impact on the reconstruction of the remnant's morphology, they severely compromise the reliability of flux-density measurements. 
For this reason, we excluded this map from the flux-density estimation of Kes~73.

 \begin{table}
        \caption{Integrated flux-density measurements of Kes~73.} 
    \centering
        \label{tab: flux densities}
        \begin{tabular}{|crl|} 
                \hline
                \hline
Freq.    & Flux density  & Reference    \\
 (GHz)   & (Jy)\quad\quad\quad &   \\

 \hline
 0.080   & $18.2\pm5.5$ & \citet{Slee_1977} \\
 0.080    & $33.0\pm18.0$ & \citet{Dulk_and_Slee_1975} \\
 0.083   & $26.0\pm10.0$ & \citet{Kovalenko_1994} \\
 0.0875  & $17\pm1.4$ & this work \\ 
 0.111   & $29.0\pm10.0$ & \citet{Kovalenko_1994}\\
 0.1185  & $15.9\pm1.3$ & this work \\  
 0.1545    & $13.1\pm1.1$ & this work \\
 0.160     & $23.0\pm2.3$ & \citet{Slee_1977}, \citet{Trushkin_1999} \\
 0.160     & $23.0\pm3.0$ & \citet{Dulk_and_Slee_1975} \\
 0.160     & $20.9\pm6.3$ & \citet{Slee_1977}, \citet{Castelletti_2021} \\
 0.2005    & $11.7\pm0.9$ & this work \\
 0.408     & $4.4\pm0.4$  & \citet{Clark_1975}\\
 0.408     & $13.0\pm2.0$ & \citet{Kassim_1989}\\
 0.408     & $12.6\pm1.9$ & \citet{Kesteven_1968}\\
 0.960     & $4.0\pm0.4$  & \citet{Trushkin_1996}\\
  1.4        & $5.5\pm0.8$  & \citet{Ingallinera_2014} \\
  2.695      & $2.85\pm0.29$  & \citet{Furst_1990} \\  
 3.9       & $1.65\pm0.17$ & \citet{Trushkin_1996}\\
 4.85      & $2.1\pm0.1$ & \citet{Griffith_1994}\\
 4.85      & $2.138\pm0.099$ & \citet{Wright_1994}\\
 5.0       & $1.9\pm0.5$ & \citet{Angerhofer_1977}\\
 5.0       & $1.4\pm0.2$ & \citet{Milne_1969}\\
 5.0          & $2.3\pm0.4$ & \citet{Ingallinera_2014} \\
 6.9       & $1.65\pm0.05$ & this work \\
 10.0       & $1.04$\tablefootmark{a} & \citet{Handa_1987} \\
 18.7      & $0.88\pm0.04$ & this work \\
 24.8      & $0.86\pm0.09$ & this work \\

 \hline 
        \end{tabular}
\tablefoot{    
\tablefoottext{a}{We reported this measurement in the SED as a reference within the frequency range covered by SRT, but we did not include it in the fit, as no uncertainty is provided in the reference paper \cite{Handa_1987}.
 } }   
\end{table}

\section{Results}
\label{sec:Results}

\begin{figure*}

    \includegraphics[width=18cm]{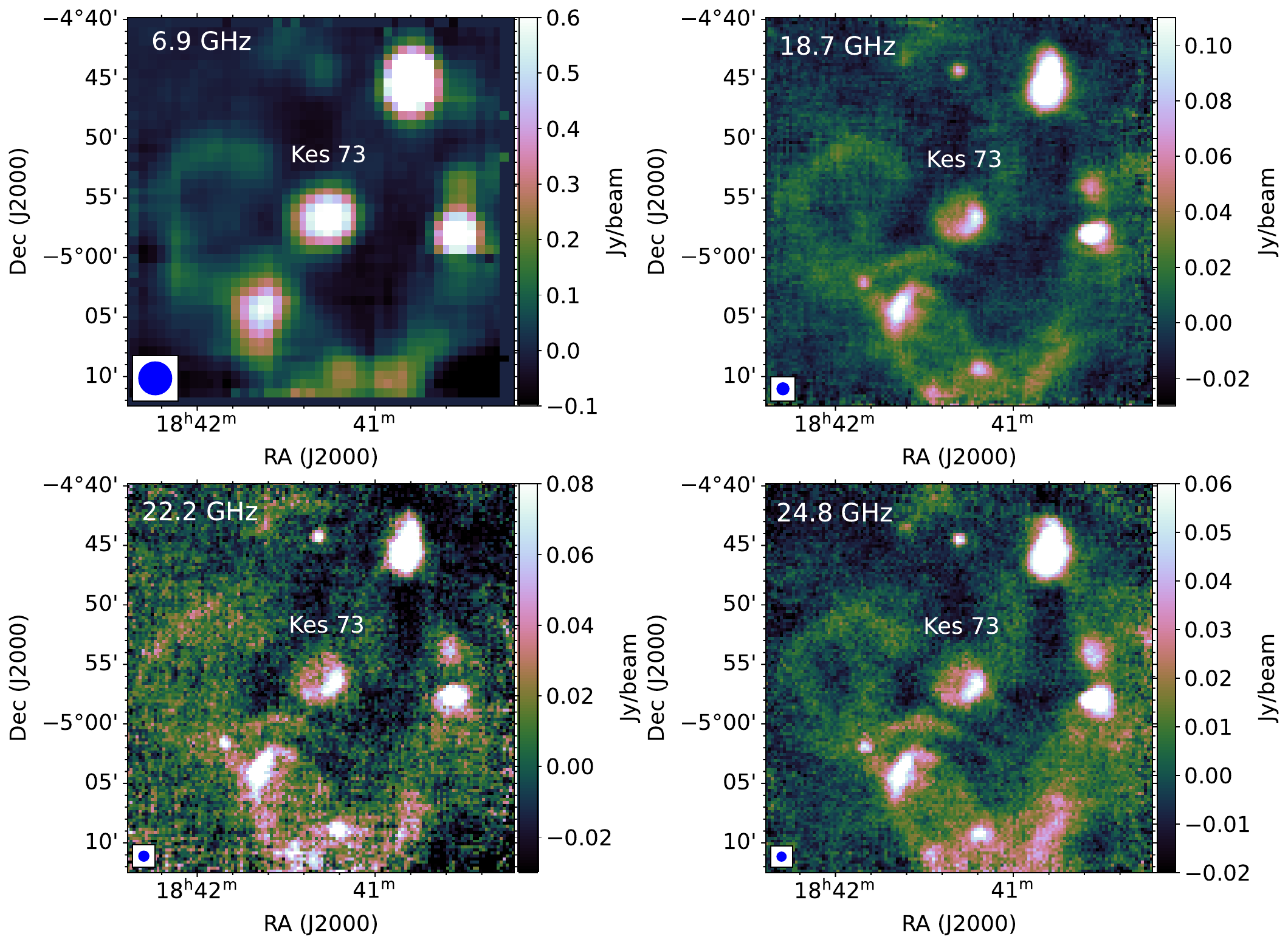}

    \caption{Continuum radio maps of Kes~73 and its surrounding field obtained with SRT at 6.9~GHz (\textit{upper left}), 18.7~GHz (\textit{upper right}), 22.2~GHz (\textit{bottom left}), and 24.8~GHz (\textit{bottom right}). 
    The pixel size is $0.7^{\prime}$ and $0.25^{\prime}$ for the $C$ band and $K$ band, respectively.
    The beam size is
    indicated by the blue circle  in the bottom-left corner of each map.}
    \label{fig: Kes73_paper_4images}
    
\end{figure*}

In Fig.~\ref{fig: Kes73_paper_4images} we present the final images obtained with the SRT at 6.9, 18.7, 22.2, and 24.8~GHz. 
Each image was produced by merging the maps obtained for each central frequency and observing session, as listed in Table~\ref{tab:obs_table}.
Among the 24.8 GHz images from different observing sessions, we show the highest-quality one that we acquired on 25 January 2019.

We calculated the flux density of Kes~73 at each frequency using aperture photometry, adopting a circular extraction region centred at ($\alpha$,$\delta$) = (18$^{\mathrm{h}}$41$^{\mathrm{m}}$17$^{\mathrm{s}}$, $-4^{\circ}56^{\prime}21^{\prime\prime}$), with a radius of $\sim 3.7^{\prime}$. 
The complex field and diffuse emission around Kes~73, clearly detected at all frequencies, make accurate baseline estimation difficult, resulting in a non-zero off-source mean flux density.
To address this, we estimated and subtracted the background contribution using a polygonal region surrounding Kes~73, and excluding unrelated sources. The resulting background flux was scaled by the ratio of the source extraction area to the background area. The final flux-density measurements are listed in Table~\ref{tab: flux densities}.

\begin{figure}

    \includegraphics[width=\columnwidth
    ]{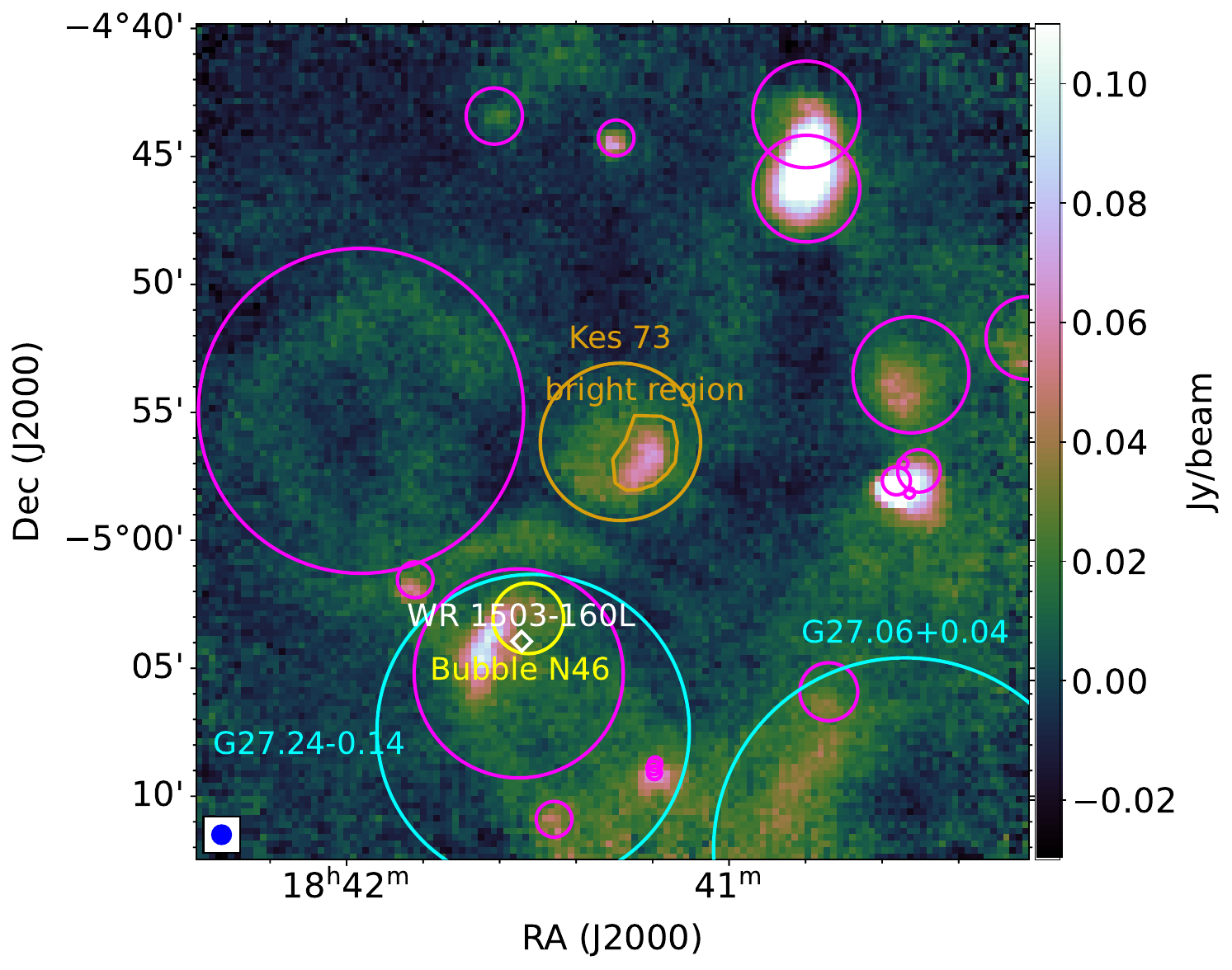}
    \label{fig:Kes73_field}

    \caption{SRT image of the Kes 73 field at 24.7~GHz. Magenta circles indicate the  H~\textsc{ii} regions catalogued in \citet{Anderson_2014}. Cyan circles correspond to two SNR candidates proposed by \citet{Anderson_2017_THOR}. The mid-IR bubble N46 \citep{Dewangan_2016} is indicated by a yellow circle, and the white diamond gives the position of the W-R star 1503-160L. We also indicate the extraction region of Kes~73 in its entirety (orange circle) and its bright region (polygon; see Sect. \ref{Sec: Multiwavelength study}).}
    \label{Kes73_field}
    
\end{figure}

The high resolution of the $K$-band SRT images allowed us to resolve the Kes~73 morphology in detail, highlighting a bright arc-like structure at the west$-$south-west edge and a more diffuse emission completing the shell-like morphology. 
The crowded field surrounding the SNR is also clearly visible in our images at all frequencies. In Fig.~\ref{Kes73_field} we highlight the radio emission from unrelated sources in the Kes~73 field at 24.8~GHz.
We clearly detect 19 H~\textsc{ii} regions, catalogued by \citet{Anderson_2014} as known (observed in both radio recombination lines and H$\alpha$ emission) and groups (WISE candidates positionally coincident with known H~\textsc{ii} region complexes).
Our image also includes part of the SNR candidate G27.06+0.04, proposed as shell-type by \citet{Anderson_2017_THOR}, and faint diffuse emission associated with the filled-centre SNR candidate G27.24$-$0.14 \citep{Anderson_2017_THOR}.
Interestingly, we resolved a bright arc-like structure just south-east of Kes~73, with no known association to any catalogued SNR or H~\textsc{ii} region.

\begin{figure}

    \includegraphics[width=\columnwidth]{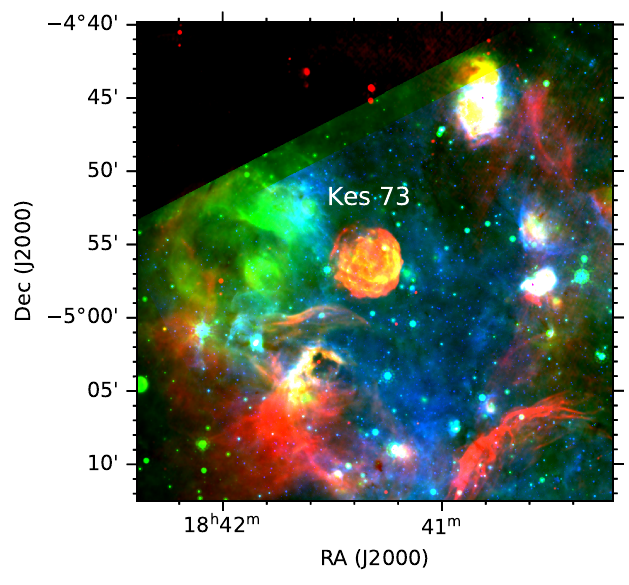}

    \caption{Three-colour image of the Kes~73 complex: SMGPS at 1.284 GHz (red), MIPSGAL at $24$~$\mu$m (resolution of $\sim6^{\prime\prime}$; green), and GLIMPSE at $8$~$\mu$m (resolution of $\sim3^{\prime\prime}$; blue).}
    \label{Fig: rgb}
    
\end{figure}

We used the high-resolution SMGPS image to perform a detailed radio–IR morphological comparison of the Kes~73 complex and investigate the nature of this arc-like feature.
The resulting three-colour image (Fig.~\ref{Fig: rgb}) shows the radio emission at 1.284~GHz in red, the mid-IR 24~$\mu$m MIPSGAL\footnote{Multiband Imaging Photometer for Spitzer Galactic plane survey} emission in green, and the 8~$\mu$m GLIMPSE emission in blue. 
The radio and 24~$\mu$m emissions from the arc structure show excellent spatial correspondence, pointing towards a thermal origin.
In Fig.~\ref{Fig: rgb}, we also note the typical radio–IR morphology of the H~\textsc{ii} regions included in the field: co-spatial radio and 24~$\mu$m emission surrounded by 8~$\mu$m emission \citep{Ingallinera_2019}.

\begin{figure}

    \includegraphics[width=\columnwidth
    ]{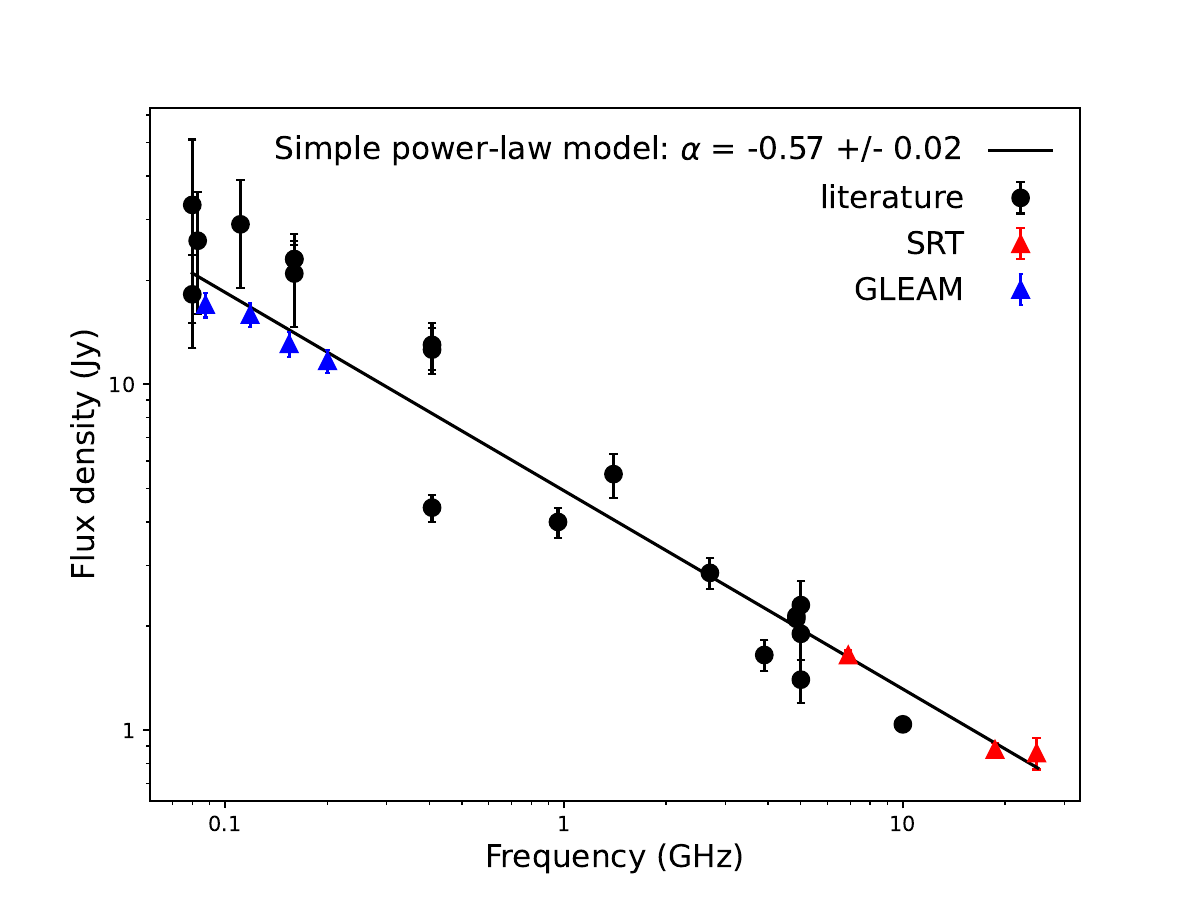}

    \caption{Integrated radio continuum spectrum of Kes~73 including our flux-density measurements from SRT images at 6.9, 18.7, and 24.8~GHz (red triangles) and the public GLEAM images (blue triangles), and flux densities available in the literature (black circles). All values and related references are reported in Table \ref{tab: flux densities}. The solid line results from the weighted least-squares fit applied to all data. The point at 10~GHz was not used for the fit.  
    }
    \label{Kes73_paper_4images}
    
\end{figure}

We studied the integrated spectrum of Kes~73 by combining the flux densities we derived from the 6.9, 18.7, and 24.8 GHz SRT images and the 0.088–0.200 GHz GLEAM images with values available in the published literature. We excluded the literature data below 0.08~GHz \citep{Castelletti_2021} to avoid bias due to free–free absorption turnover. We also excluded the 22.2 GHz SRT image from the integrated flux analysis due to its lower quality  (Fig.~\ref{Kes73_paper_4images}) and high rms value (Table~\ref{tab:obs_table}). Both factors are attributable to suboptimal and highly variable weather conditions during the observing session.  Among the 24.8 GHz images, we used the one from 25 January 2019 (Fig.~\ref{Kes73_paper_4images}, bottom right), which has the lowest rms (Table~\ref{tab:obs_table}). We included the literature measurement at $10$~GHz in the spectral energy distribution (SED) as a useful reference point within the frequency range covered by the SRT, although we excluded it from the fit due to the absence of uncertainty estimates in the reference paper \citep{Handa_1987}. All flux densities are listed in Table~\ref{tab: flux densities} and plotted in Fig. \ref{Kes73_paper_4images}. By fitting the data with a simple synchrotron power-law model ($S_{\nu} \propto \nu^{\alpha}$), we derived an integrated spectral index of $\alpha = -0.57 \pm 0.02$, significantly higher than the value $\alpha = -0.690 \pm 0.035$ reported by \citet{Castelletti_2021} in the 0.074–5.0 GHz range. 
We note that the literature fluxes exhibit considerable scatter, whereas GLEAM and SRT data are consistent across a wide frequency range, including the highest ever observed for this source that dominates the spectral fit.

\begin{figure}
    \centering
    \includegraphics[width=\columnwidth]{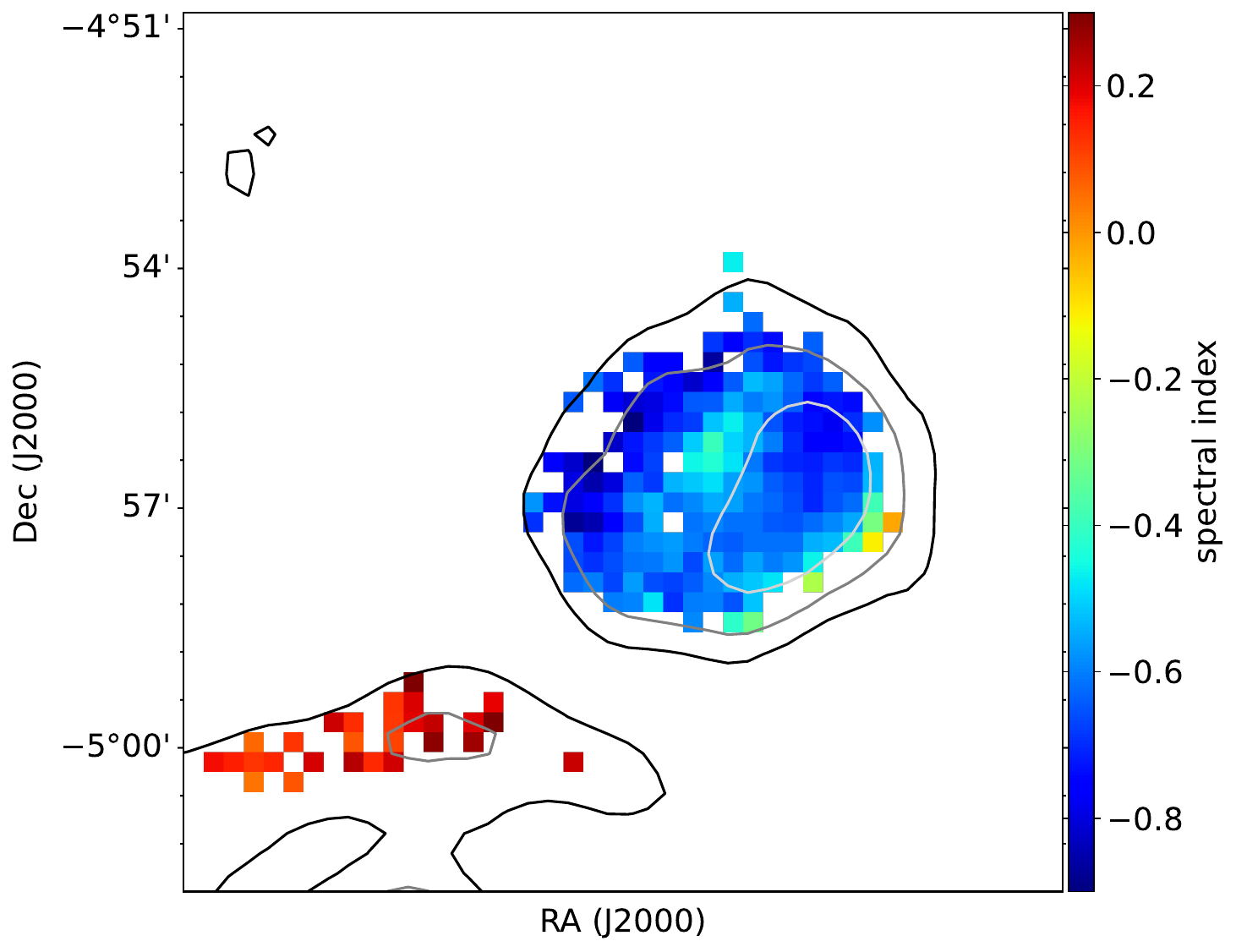}
    \includegraphics[width=\columnwidth]{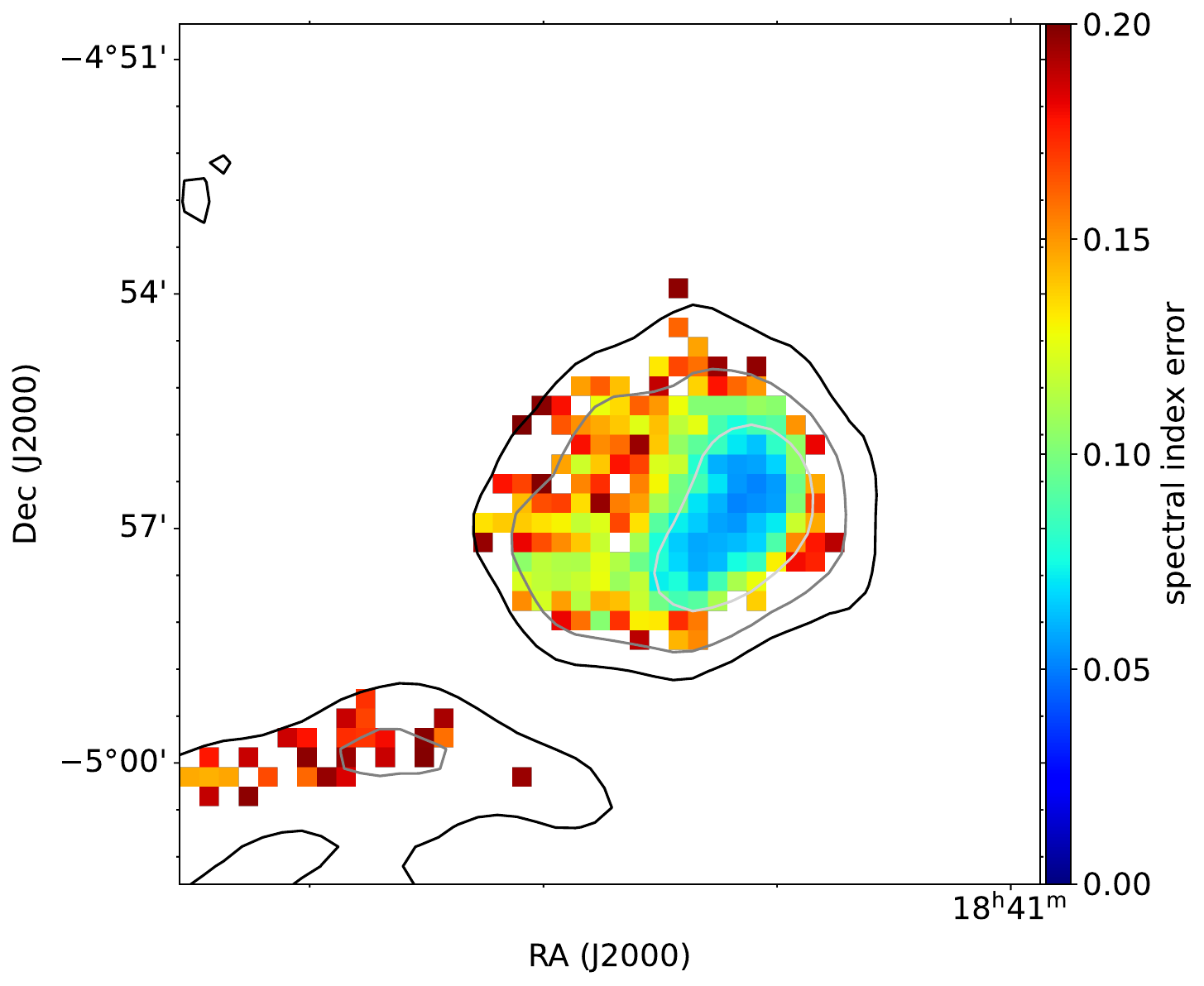}    
    \caption{Spectral-index map of Kes~73 (\textit{top}) and related error map (\textit{bottom}) produced from SRT and VLA+GBT data in the 1.4-24.8 GHz frequency range. The contours indicate the intensity levels at 0.009, 0.018, and 0.035~Jy beam$^{-1}$ of the 24.8 GHz SRT map of Kes~73.}
    \label{fig:spix}
\end{figure}

To study spatial variations in the spectral index across the remnant, we produced a spectral-index map using the 24.8 GHz SRT and 1.4 GHz interferometric images from \citet{Ingallinera_2014} with a resolution of $18^{\prime\prime} \times 12^{\prime\prime}$. 
The latter was obtained by combining Green Bank Telescope (GBT) and \textit{Karl G. Jansky }Very Large Array (VLA) data.
Starting from the background subtracted maps, we convolved and re-gridded the 1.4 GHz map to match the beam size, coordinate system, and pixel scale of the 24.8 GHz SRT image using the \textsc{casa} tasks \textsc{convolve2d} and \textsc{imregrid}, respectively. Spectral index errors were computed via standard error propagation, and the final spectral-index map was masked where the uncertainty exceeds 0.2, ensuring reliable values. 
The resulting spectral-index map and its corresponding error map are shown in Fig.~\ref{fig:spix}.
The map shows different spectral regions: a bright western region with a mean spectral index of $\alpha= -0.59 \pm 0.06$; the easternmost part of the shell with $\alpha = -0.73 \pm 0.04$; and a central region with a flatter spectrum ($\alpha = -0.53 \pm 0.05$), corresponding to the low-brightness feature visible in the 1.4 GHz map (Fig.~5, left panel, in \citealt{Ingallinera_2014}). Although these values are consistent within $3\sigma$, they confirm an asymmetric spectral trend, with the western region exhibiting a flatter index than the eastern one.

\section{Multi-wavelength study}
\label{Sec:5}
\subsection{Modelling the non-thermal emission}
\label{Sec: Multiwavelength study}

Radio and $\gamma$-ray emissions are crucial windows for studying the energetics of CRs accelerated in SNRs. 
While hadronic processes contribute to $\gamma$-ray emission through $\pi^0$ decay following proton–proton interactions, leptonic processes involve both bands: synchrotron emission, observable in the radio band, and inverse Compton (IC) and bremsstrahlung processes, which contribute to high-energy emission.
For this reason, observational radio parameters (e.g. integrated spectral index, spectral cutoff) are essential to constrain the electronic contribution to the $\gamma$-ray band and to properly understand the leptonic and/or hadronic nature of the SNR emission.

From a multi-wavelength perspective, SNRs interacting with MCs are of particular interest, since they are sites of enhanced CR production. In these cases, we often observe a brighter and spectrally flatter region in the radio band, accompanied by a matching $\gamma$-ray source that traces the enhanced proton collisions with the ambient gas density \citep{Dubner_2015}. The possible interaction of the western shell of Kes~73 with nearby MCs has been suggested based on the widened line profiles of CO emission  (\citealt{Kilpatrick_2016}, \citealt{Liu_2017}) and the detection of near-IR [Fe II] and H$_2$ lines \citep{Lee_2019}. 
A recent radio study also attributed the observed free–free absorption to ionized gas produced by the interaction between the supernova shock and the ISM \citep{Castelletti_2021}. A $\gamma$-ray counterpart was also detected by HESS\footnote{High Energy Stereoscopic System} (HESS J1841–055, \citealt{Aharonian_2008}), \textit{Fermi}-LAT (source A in \citealt{Liu_2017}), and MAGIC\footnote{Major Atmospheric Gamma Imaging Cherenkov Telescope} \citep{MAGIC_2020}.

High-resolution radio images allow us to identify the remnant structures closely involved in the SNR–MC interaction and to focus the spectral analysis specifically on those regions.
In this way, we can obtain the radio parameters actually associated with the electron population produced in the same region from which the $\gamma$-ray emission arises, and use them to more tightly constrain the overall SNR non-thermal emission. 
We should note that the radio spectral index value strongly affects the non-thermal models applied to radio and $\gamma$-ray data: the more accurate the estimate, the more reliable the inferred particle energetics. Nevertheless, nearly all multi-wavelength studies conducted on SNRs so far rely on integrated radio flux densities over the whole remnant, due to the difficulty of carrying out high-resolution maps across a wide frequency range.  In many cases, however, the integrated spectral index is not representative of the different electron populations arising in specific SNR regions, particularly where SNR–MC interaction is taking place \citep{Egron_2017, Loru_2024}.  A significant spectral variation across the remnant was also observed in Kes~73 between 1.4 and 5~GHz, ranging from $\sim -0.8$ in the fainter eastern region to $\sim -0.2$ in the brightest western one (Fig. 6 in \citealt{Ingallinera_2014}). We observed the same trend at higher frequencies by comparing our high-resolution SRT image at 24.8~GHz with the 1.4 GHz VLA+GBT map from \citet[see our Fig.~\ref{fig:spix}]{Ingallinera_2014}, although with lower significance.

\begin{figure}

    \includegraphics[width=\columnwidth
    ]{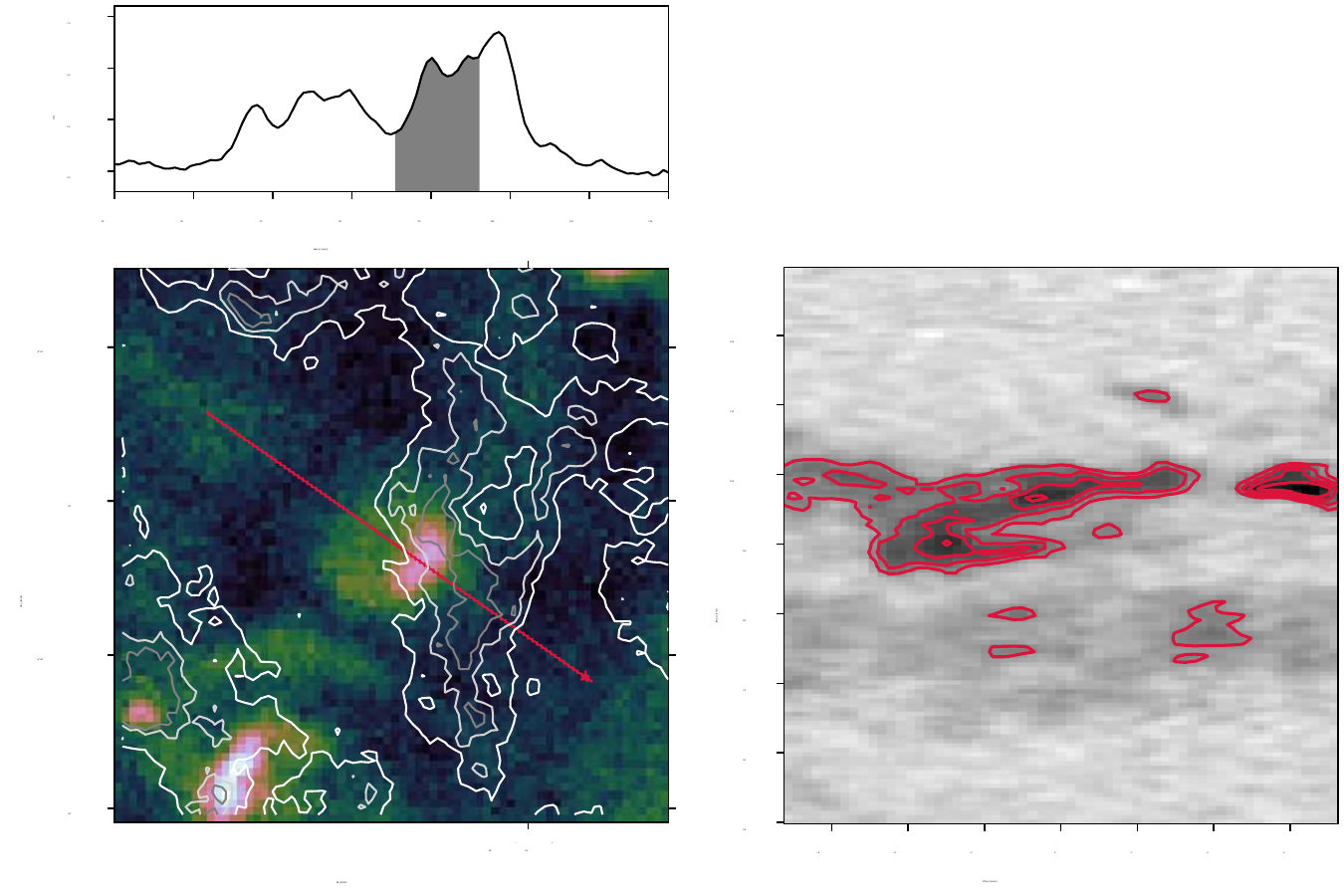}

    \caption{Velocity-spatial features and $^{12}\mathrm{CO}$ spectrum of the MC co-located with Kes~73.
    \textit{Upper}: $^{12}\mathrm{CO}$ spectrum from the extraction region of Kes~73 (orange circle in Fig. \ref{Kes73_paper_radio_CO_70_100_120K_gamma}). The grey-shaded region  highlights the $85-96$~km~s$^{-1}$ velocity range, corresponding to the MC material that overlaps the remnant. \textit{Bottom left}: 24.8 GHz SRT image of Kes~73 (same as in Fig. \ref{fig: Kes73_paper_4images}, bottom right)
    with the 70, 100, and 120~K temperature levels referred to the $^{12}\mathrm{CO}$ emission integrated over the range $85-96$~km~s$^{-1}$. \textit{Bottom right}: PV diagram of the $85-96$~km~s$^{-1}$ $^{12}\mathrm{CO}$ material across the direction indicated by the red arrow. Red contours indicate the temperature levels at 5, 7, 9, and 11~K.
    }
    \label{profilo_velocità_Kes_reg}
    
\end{figure}

\begin{figure}

    \includegraphics[width=\columnwidth
    ]{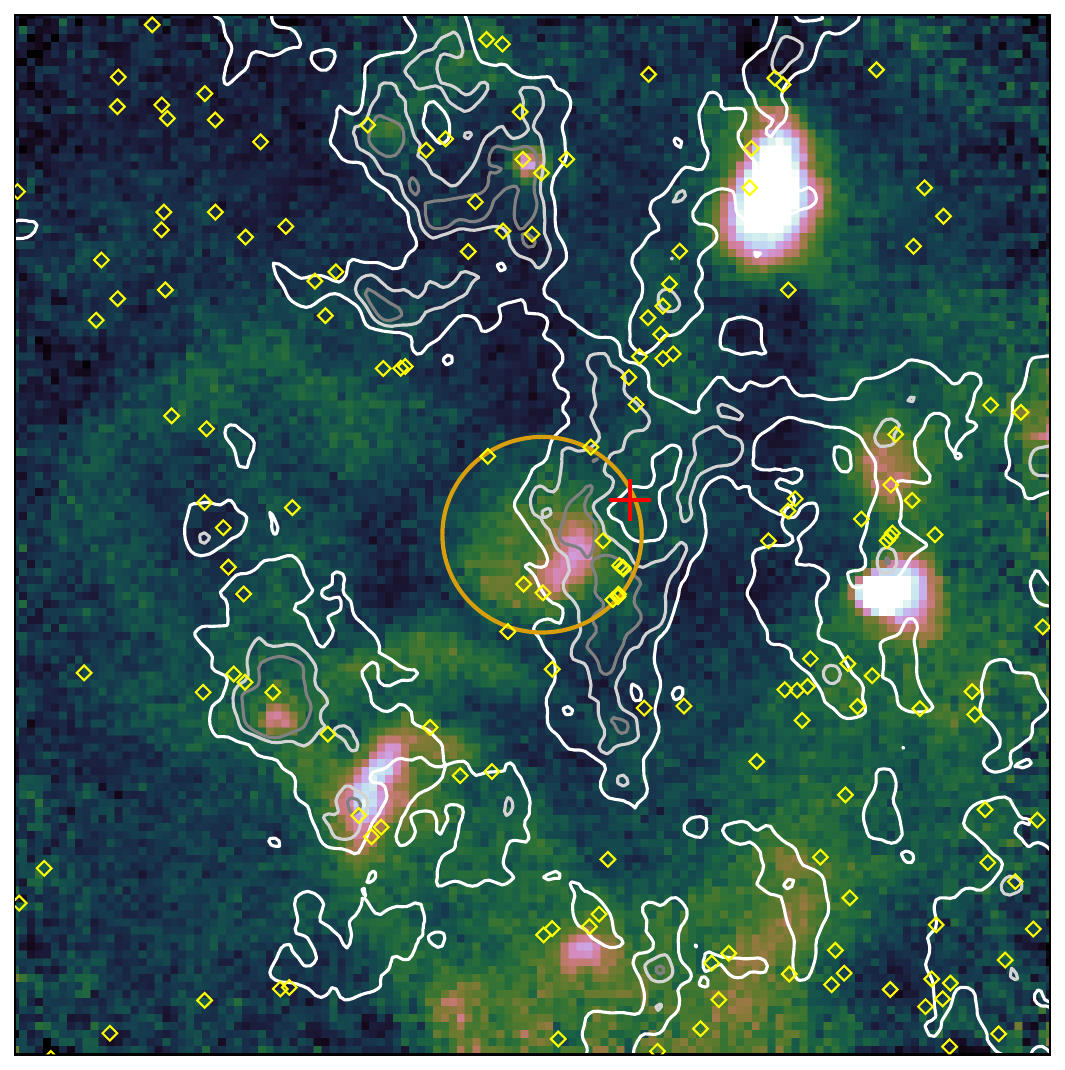}

    \caption{SRT radio image of Kes~73 at 24.8~GHz with contour levels of the $^{12}\mathrm{CO}$ emission superposed. The $^{12}\mathrm{CO}$ emission is integrated on the $85$~$-$~$96$~km~s$^{-1}$ velocity interval, and the white, light grey, and grey contours correspond to the temperature levels at 70, 100, and 120~K, respectively. The orange circle marks the extraction region we used to estimate the whole  flux density of Kes~73 on our $K$-band maps, and the red cross the centroid of the $Fermi$-LAT $\gamma$-ray source. The yellow diamonds indicate the YSOs from  the SPICY catalogue.
}
    \label{Kes73_paper_radio_CO_70_100_120K_gamma}
    
\end{figure}

We used the CO-line data from the FOREST Unbiased Galactic Plane Imaging (FUGIN) survey\footnote{\url{http://jvo.nao.ac.jp/portal/nobeyama/fugin.do}} (resolution of $\sim20^{\prime\prime}$, \citealt{Umemoto_2017}), based on observations from the Nobeyama Radio Observatory, to investigate the molecular environment around Kes~73 and strictly constrain the radio region associated with the MC interaction. We compared the 24.8 GHz radio map with the FUGIN $^{12}$CO ($J=1$–0) emission, integrated over a velocity range of 85–96~km\,s$^{-1}$, coinciding with the most prominent peak in the spectral profile extracted from the Kes~73 region centred at ($\alpha$, $\delta$) = (18$^{\mathrm{h}}$41$^{\mathrm{m}}$18.0559$^{\mathrm{s}}$, $-4^{\circ}56^{\prime}24.313^{\prime\prime}$),
with a radius of $\sim 3.1^{\prime}$ (see Fig.~\ref{profilo_velocità_Kes_reg}, upper plot). As shown in Fig.~\ref{Kes73_paper_radio_CO_70_100_120K_gamma}, the $^{12}$CO emission overlaps the western side of the Kes~73 extraction region (indicated by the orange circle) and appears morphologically coincident with its brightest radio region. 
By a red cross, we marked the centroid of the extended GeV \textit{Fermi}-LAT source identified by \citet{Liu_2017} as associated with Kes~73 (referred to as source A). 
The young stellar objects (YSOs) from the \textit{Spitzer}/IRAC Candidate YSO (SPICY) catalogue\footnote{\href{https://irsa.ipac.caltech.edu/data/SPITZER/SPICY/overview.html}{https://irsa.ipac.caltech.edu/data/SPITZER/SPICY/overview.html}} that fall within the Kes~73 field are indicated by yellow diamonds in Fig.~\ref{Kes73_paper_radio_CO_70_100_120K_gamma}. 
A spatial correlation is observed between their distribution and the $^{12}$CO emission, particularly along the western edge of the remnant and in the northern molecular regions, possibly indicating enhanced star-forming activity.

We assumed the updated solar motion parameters from \citet[Galactocentric radius of the Sun $R_0 = 8.34$~kpc and the orbital velocity of the Sun around the Galactic centre $V_0 = 241.0$~km\,s$^{-1}$]{Reid_2014} to estimate the near and far kinematic distances of the MC using the Galactic rotation model of \citet{Reid_2014}\footnote{We used the \href{https://www.treywenger.com/kd/}{Kinematic Distance Calculation Tool} based on \citet{2018Wenger_}.}.
By considering the peak of the $^{12}\mathrm{CO}$ emission at a velocity of $\sim 90$~km\,s$^{-1}$, located at ($l$, $b$) = ($27.3811581^{\circ}$, $-0.0034697^{\circ}$), we obtained a kinematic distance of either 5.37~kpc or 9.44~kpc. The far distance falls within the range of 7.5–9.8~kpc derived for Kes~73 by \citet{Tian_2008} based on an H~\textsc{i} absorption study. 
However, \citet{Ranasinghe_2018} pointed out that the absorption features used by \citet{Tian_2008} to estimate this range are most likely spurious. 
They also revised the distance of Kes~73 to 5.8~kpc using a new H~\textsc{i} absorption analysis. 
This kinematic distance has since been confirmed by a high-resolution near-IR spectroscopic study \citep{Lee_2020}. 
The value of 5.8~kpc is consistent with our near-distance estimate, and we adopt it as the distance to the SNR–MC interaction system based on the most recent literature.
To further constrain the interaction between Kes~73 and the co-distant MC, 
we searched for systematic velocity gradients in the position–velocity (PV) diagram, which trace variations in the MC velocity (either approaching or receding along the line of sight) caused by the SNR shock \citep{Feng_2024}.
We used the GREG graphical utility package of GILDAS\footnote{\url{https://www.iram.fr/IRAMFR/GILDAS/}} 
to produce the PV diagram of the $^{12}$CO emission shown in Fig.~\ref{profilo_velocità_Kes_reg} (bottom right), along the direction indicated by the arrow (bottom left). The PV diagram reveals an arc-like structure spanning from $\sim 86$ to $\sim 100$~km\,s$^{-1}$, suggesting a strong shock pushing the cloud away along the line of sight. This kinematic signature provides conclusive evidence of SNR–MC interaction in the bright western radio region of Kes~73. 

\begin{figure}

    \includegraphics[width=\columnwidth]{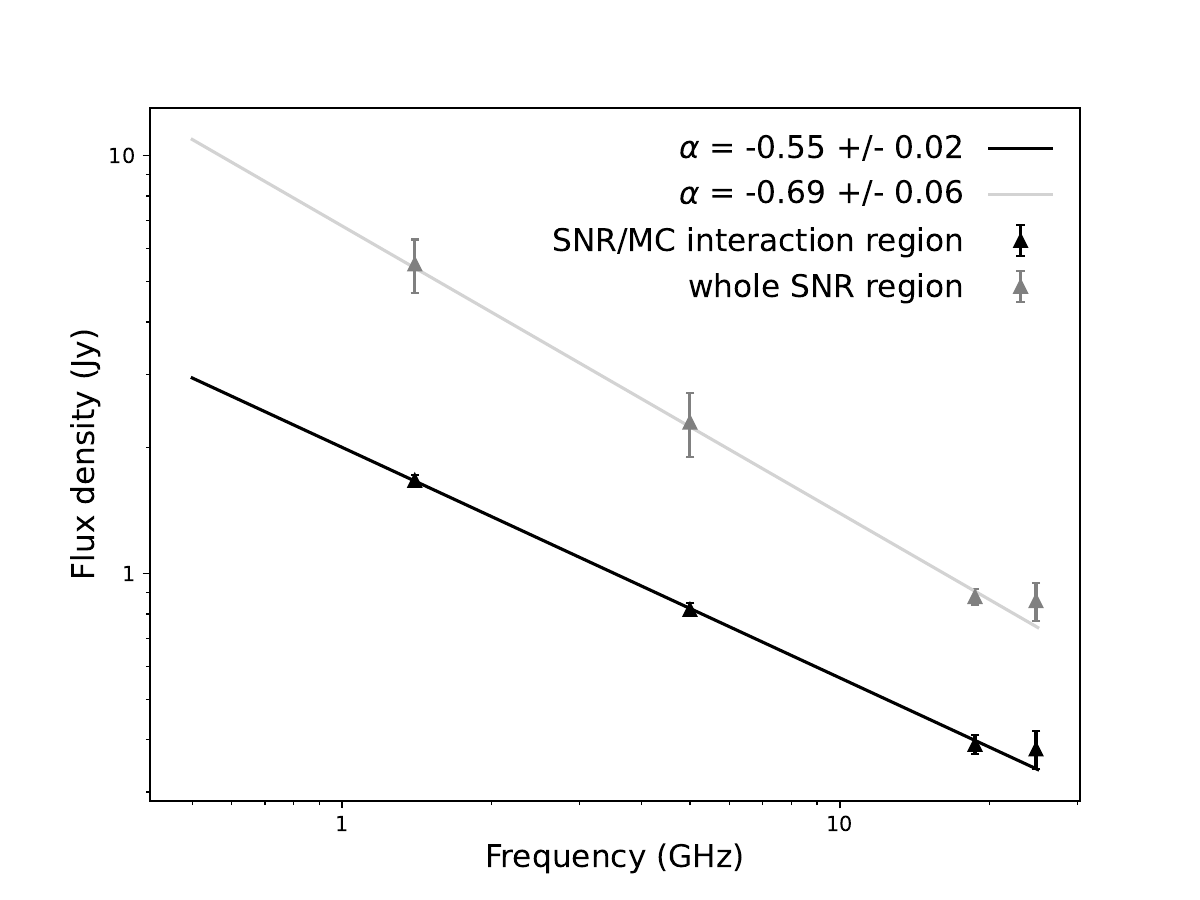}

    \caption{Integrated spectrum of the SNR-MC interaction region (`bright region' in Fig. \ref{fig:Kes73_field}) of Kes~73 from 1.4 and 5.0 GHz VLA+GBT and 18.7 and 24.8 GHz SRT images and the related fit from a simple power-law model. 
    For comparison, we also plot the integrated flux densities calculated by considering the whole Kes~73 extension (orange circle in Fig. \ref{fig:Kes73_field}) and related fit.
    }
    \label{VLA_SRT spectrum}
    
\end{figure}

By comparing the radio and $^{12}$CO emissions shown in Fig.~\ref{Kes73_paper_radio_CO_70_100_120K_gamma}, we identified the SNR–MC interaction region as coinciding with the brightest portion of the Kes~73 shell.
Based on this, we used the VLA+GBT and SRT images to calculate the integrated flux densities at 1.4, 5, 18.7 and 24.8~GHz for both the interaction region (orange polygon labelled `bright region' in Fig. \ref{Kes73_field}) and the entire SNR (the extraction region, including all the remnant regions, which is indicated by the orange circle in Fig.~\ref{Kes73_field}). As shown in Fig. \ref{VLA_SRT spectrum}, the resulting spectral index for the interaction region ($\alpha = -0.55 \pm 0.02$) is significantly higher than that of the whole SNR calculated only on these four frequencies ($\alpha = -0.69 \pm 0.06$).

We used these radio results, together with the \textit{Fermi}-LAT flux measurements for source A reported by \citet[Table 4]{Liu_2017}, to model the SED of Kes~73 with the \textsc{NAIMA} Python package \citep{Zabalza_2015} from radio to GeV energies, including both leptonic and hadronic components resulting from the SNR–MC interaction. We assumed a single population of electrons for this region, and modelled the particle energy distributions (for both electrons and protons) using a power law with an exponential cutoff:

\begin{equation}
    f(E_i) = A_i \left( \frac{E}{E_i} \right)^{-s_i} \exp\left( -\frac{E}{E_{c,i}} \right)
,\end{equation}

\noindent where $i = e, p$ refers to electrons and protons, respectively. 
$A_i$ is the normalization factor, $E_i$ the reference energy (fixed at 1~TeV), $E_{c,i}$ the cutoff energy, and $s_i$ the spectral index,  related to the radio spectral index via $s_e = 1-2\alpha $.
The electron-to-proton ratio at 1~TeV is described by $K_{ep} = A_e / A_p$ in the lepto-hadronic scenario. 

\begin{figure*}
        \includegraphics[width=\columnwidth
    ]{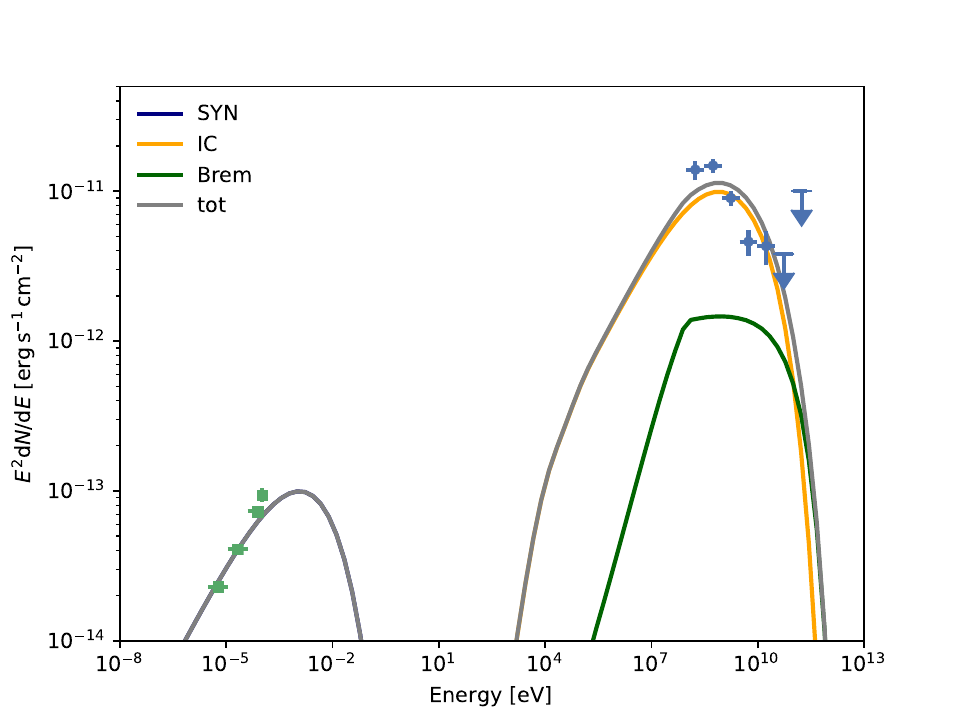}
        \includegraphics[width=\columnwidth
    ]{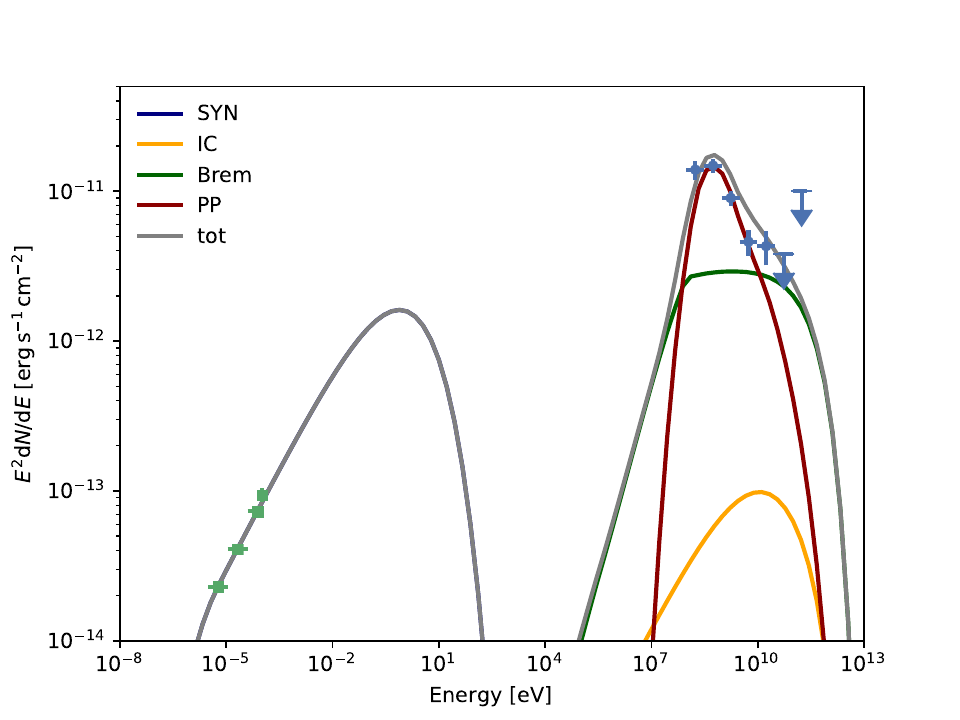}
    
    \caption{Multi-wavelength SED of the Kes~73-MC interaction region (`bright region' in Fig.\ref{fig:Kes73_field}) in a
    pure leptonic (\textit{left}) and lepto-hadronic scenario (\textit{right}). The model parameters are listed in Table \ref{tab: fit models results}. Blue, yellow, and green lines indicate the synchrotron, IC, and bremsstrahlung emission contribution from the same electron distribution, respectively. The red line indicates the proton-proton emission from CR protons.}
    \label{pure leptonic model SED}
\end{figure*}

We first applied a pure leptonic model including synchrotron, IC, and bremsstrahlung contributions. 
The radio dataset includes flux densities from the VLA+GBT maps at 1.4 and 5~GHz, and the SRT maps at 18.7 and 24.8~GHz, 
all extracted from the bright region. For the $\gamma$-ray band, we used the fluxes of source A from \citet{Liu_2017}. 
We assumed IC scattering of relativistic electrons on the cosmic microwave background (2.72~K, 0.261~eV\,cm$^{-3}$) and far-IR dust photons (26.5~K, 0.415~eV\,cm$^{-3}$), 
adopting the default values from the \textsc{NAIMA} package. We fixed the SNR–MC system distance to 5.8~kpc, following \citet{Lee_2020}, and used a power-law index of $s_e = 2.1$, 
consistent with the radio spectral index $\alpha = -0.55$ derived for the bright region. The resulting fit (see Fig.~\ref{pure leptonic model SED}, and parameters in Table~\ref{tab: fit models results}) 
shows that the bremsstrahlung contribution is negligible compared to IC.
However, the model fails to reproduce the observed synchrotron emission, yielding a cutoff energy of $E_{c,e} = 0.27 \pm 0.03$~TeV. 
Assuming that an electron with energy $E$ accelerated in a magnetic field $B$ radiates its peak power at a characteristic frequency $\nu_c$ according to the relation \citep{Reynolds_2008}
\begin{equation*}
    E = 14.7 \left( \frac{\nu_c/\mathrm{GHz}}{B/\mathrm{\mu G}}\right)^{\frac{1}{2}} \mathrm{GeV,} \qquad
    \label{Obs eq.}
\end{equation*}
the derived values of $E_{c,e}$ and $B = 0.52~\mu$G imply a synchrotron cutoff frequency of $\nu_c \sim 175$~GHz ($\sim 7.2 \times 10^{-4}$~eV), which is clearly underestimated compared to our data.
This result also implies that $B = 0.52~\mu$G is a lower limit. Furthermore, the pure leptonic model requires a total electron energy of $9.6 \times 10^{49}$~erg above 1 GHz, a too high value for the electron energy budget. Indeed, assuming 10\% of the supernova kinetic energy ($E_{\mathrm{SN}} = 10^{51}$~erg) is transferred to accelerated CRs, and 1\% of that to electrons, or $\sim 10^{48}$~erg \citep{SNR_book_2020}.

The lepto-hadronic model better reproduces the radio spectrum, predicting a synchrotron cutoff $\nu_c \sim 1.5 \times 10^5$~GHz ($\sim 0.6$~eV), 
far above our frequency range. The $\gamma$-ray spectrum is well fitted by pion-decay emission, with bremsstrahlung becoming significant above $\sim 10$~GeV.
The model yields electron and proton energy budgets of $\sim 0.02$\% and $1.8$\% of the explosion energy, respectively, which are consistent with theoretical expectations. We obtained an ion density of the circumstellar gas of $130^{+90}_{-50}$~cm$^{-3}$, matching the estimated electron density  ($\sim 70$–$110$~cm$^{-3}$) from \citet{Castelletti_2021}. The inferred magnetic field strength of $25$~$\mu$G lies within the range of expected values by SNR shock compression of the
ISM magnetic field ($B \sim 10$–$100~\mu$G, \citealt{Urosevic_2014}).

At $\nu \gtrsim 30$~GHz, both non-thermal synchrotron and thermal emission (from dust and gas) can contribute to the SNR spectrum. 
The observed IR thermal emission from the shocked ISM  provides
a constraint on the highest synchrotron flux densities that our model can predict at these frequencies. 
With this aim, we verified that the IR flux density of $16.6\pm4.5$~Jy ($9.96 \times 10^{-11}$~erg~cm$^{-2}$~s$^{-1})$ at $600$~GHz (500~µm or $2.5 \times 10{-3}$~eV) calculated by \citet{Priestley_2021}
is well above the modelled synchrotron emission, and therefore not in conflict with our lepto-hadronic model.

Finally, comparison via the Bayesian information criterion (BIC; \citealt{Schwarz_1978}) yields 
$\Delta \mathrm{BIC} = \mathrm{BIC}_{\mathrm{leptonic}} - \mathrm{BIC}_{\mathrm{lepto-hadronic}} > 6$. This is clearly evidence in favour of the lepto-hadronic scenario.

For completeness, we also tested both models using the far kinematic distance of 9.44~kpc, which we derived for the MC co-located with the western boundary of Kes~73 and which is compatible with the distance estimated by \citet{Tian_2008}. We found that changing the distance does not significantly affect most parameters: the magnetic field, ion density, and the energy cutoffs for protons and electrons remain consistent within the uncertainties. The fits obtained using the two distances are also comparable for the two models, which yield the same BIC value. The main difference concerns the total energy, which increases by about a factor of three compared to the results obtained assuming a distance of 5.8~kpc. In the pure leptonic model, we obtained $W_e$ ($E > 1$~GeV) = $(2.7^{+0.6}_{-0.4}) \times 10^{50}$~erg, while in the lepto-hadronic scenario we derived $W_e = (6^{+3}_{-2}) \times 10^{47}$~erg and $W_p = (4.5^{+5}_{-2}) \times 10^{49}$~erg. Despite these higher values, the resulting energies still fall within the expected range for these objects. We therefore conclude that our model does not provide a clear indication of which distance is more reliable. Nevertheless, we note that the distance of 5.8~kpc recently updated by \citet{Lee_2020} is widely adopted in the current literature, which is why we used it as our reference value.

 \begin{table}
\noindent
        \centering
        \caption{Resulting parameters for the pure leptonic and the hybrid lepto-hadronic models of the Kes 73-MC interaction region. } 
        \label{tab: fit models results}
        \begin{tabular}{|l|c|l|} 
                \hline
                \hline
Parameters       & leptonic & lepto-hadronic    \\

 \hline

 A$_e$  (TeV~$^{-1}$) & $(7.7^{+1.4}_{-1.1}) \times 10^{48}$ & $(1.6^{+1.1}_{-0.8}) \times 10^{46}$*\\
 A$_p$  (TeV~$^{-1}$) & $-$ & $ (5.8^{+3.0}_{-1.8}) \times 10^{46}$ \\
 $s_{e}$ & 2.1 & 2.1\\
 $E_{c_{e}}$ (TeV) & $0.27\pm0.03$ & $1.1\pm0.1$\\
 $s_{p}$ & $-$ & $2.7\pm0.1$ \\
 $E_{c_{p}}$ (TeV) & $-$ & $0.995\pm0.007$ \\
 B ($\mu$~G) & $0.52\pm0.05$ & $25\pm6$ \\
 n$_H$ (cm~$^{-3}$)      & $0.3^{+0.6}_{-0.2}$ & $130^{+90}_{-50}$ \\
 K$_{ep}$ & $-$ & $0.28\pm0.05$ \\
 W$_{e}$ (E$>1$~GeV) (erg) & $(9.6^{+1.4}_{-1.4}) \times 10^{49}$ & $(2.4^{+1.5}_{-0.8}) \times 10^{47}$  \\
 W$_{p}$ (E$>1$~GeV) (erg)& $-$ & $ (1.8^{+1.5}_{-0.8}) \times 10^{49}$  \\
  BIC & 58 & 29 \\

\hline 

        \end{tabular}
        \tablefoot{We indicated with A$_e$ and A$_p$ the amplitude of the electron and proton energy distribution, respectively. 
    $s_{e}$ and $E_{c_{e}}$ are the slope and the energy cutoff of the electron energy distribution, while $s_{p}$ and $E_{c_{p}}$ are the same parameters but referred to the proton energy distribution.  W$_{e}$ and W$_{p}$ are the total energy of the CR electrons and protons (calculated for particle energies above 1~GeV), respectively.
    B is the magnetic filed, n$_H$ the ion density of the target gas and  K$_{ep}$ is defined as $\frac{A_e}{A_p}$ and describes the ratio of electrons with respect to protons at the reference energy $E_0$ that we set to 1~TeV. We indicated with BIC the Bayesian information criterion parameter defined as BIC$= k \log(n) - 2 \log(L_{best})$, 
    where $k$ is the number of parameters in the model, $n$ the number of data points, and $L_{best}$ the likelihood associated with the best-fit parameters.}
{\raggedright  *The amplitude of the electron energy distribution is given by the model as $A_e = A_p \times K_{ep}$.
 \par}
\end{table}

\subsection{The crowded Kes~73 field}
\label{sec:ThecrowedKes73field}

Core-collapse SNRs result from explosions of massive stars that are born and evolve within star clusters or OB associations, where star formation activity remains high over long timescales \citep{Dubner_2015}. 
A spatial correlation between core-collapse SNRs and star-forming regions is therefore expected, and has been observed for several objects (e.g. W30, G54.1+0.3, G357.7+0.3, IC443, and G18.8+0.3; \citealt{Dubner_2015}, and references therein). Since star formation tends to occur in high-density regions of the Galaxy, often within or near MCs, most core-collapse SNRs are expected to lie in the proximity to MCs, with the possibility of interacting with them.
Nevertheless, the physical association between SNR–MC interacting systems and nearby star-forming regions does not necessarily imply a causal relationship \citep{Dubner_2015}. In particular, the hypothesis that star formation can be triggered by SNR–ISM interaction remains under debate.  
Ongoing observational and theoretical studies of an increasing number of such SNRs are aimed at constraining the mutual influence between SNR shocks and star-forming activity. 

It is important to consider that, massive stars strongly modify their immediate environments during their evolution so that core-collapse SNRs are generally expected to interact with the dense material surrounding the stellar wind bubble during their later evolutionary stages. This interaction is likely to occur earlier in cases where the progenitor star has created a relatively small wind bubble, such as for stars with lower initial masses (e.g. $\sim$12~$M_\odot$), compared to the more extended cavities formed by very massive progenitors \citep{SNR_book_2020}.

Kes~73 likely originated from a core-collapse supernova (Type II or Ib), with a progenitor mass of $\sim$12~$M_\odot$ \citep{Zohu_2019}. 
It is located within a complex environment that includes the star-forming region N46, the Wolf–Rayet (W-R) star 1503-160L, the two SNR candidates G27.06+0.04 and G27.24-0.14, and several H~\textsc{ii} regions and YSOs. 
A detailed radio view of the crowded field surrounding Kes~73 is provided by the high-resolution SMGPS image shown in Fig. \ref{Fig: rgb}. 
This scenario, along with the evidence of interaction with the nearby $^{12}$CO MC discussed in Sect.~\ref{Sec: Multiwavelength study}, 
further supports Kes~73 as a promising candidate for a possible association between SNR evolution and star-forming activity.
Though \citet{Dewangan_2016} dismissed the possibility that the nearby H~\textsc{ii} regions and W-R are physically related to Kes~73, the revised distance of $5.8$~kpc\footnote{for comparison, \citet{Dewangan_2016} used a distance of $7.5-9.1$~kpc for Kes~73.}, which we are adopting in this work, places the SNR much closer to the W-R 1503-160L ($5.2 \pm 0.7$~kpc) and the bubble N46 ($4.8 \pm 1.4$~kpc). It is therefore possible that many of these sources share a common origin.

\section{Conclusions}
\label{sec:Conclusions}

We have presented new, sensitive single-dish radio-continuum images of the SNR Kes~73, obtained with SRT between $6.9$ and $24.8$~GHz.
These observations allowed us to investigate the integrated radio spectrum of Kes~73 in a previously unexplored high-frequency domain, obtaining a spectral index of $\alpha= -0.57\pm0.02$ and effectively ruling out any spectral curvature up to these frequencies. This behaviour is consistent with expectations for young SNRs still undergoing efficient particle acceleration \citep{Urosevic_2014}. 

We also measured the flux densities in the $0.088$–$0.200$ GHz range using public GLEAM survey images, thereby extending the frequency span and improving the accuracy of the integrated spectral-index determination. 
Furthermore, we exploited the high resolution ($\sim0.25^{\prime}$) of the SRT $K$-band maps to investigate the high-frequency morphology and spatially resolved spectral properties of this small SNR. 
The analysis reveals an asymmetric shell structure, with the flattest-spectrum and brightest region located at the western boundary.
By comparing the $24.8$ GHz SRT image with $^{12}$CO-line data from the FUGIN survey, we found a consistent velocity between Kes~73 and the molecular emission at $\sim 90$~km\,s$^{-1}$ (which is compatible with either the near or far kinematic distance).
This emission shows a morphological coincidence with the bright western radio feature of the remnant. 
We also observed a velocity gradient between $\sim 86$ and $100$~km\,s$^{-1}$ in the PV diagram of the $^{12}$CO emission co-located with the bright radio region,
providing further confirmation of the SNR–MC interaction. 

The presence of an extended \textit{Fermi}-LAT $\gamma$-ray source with its centroid located next to the brightest radio region 
provides strong evidence of enhanced CR production. 
We used the spectrum between 1.4 and 24.8~GHz of the radio-bright, flat-spectrum region to model the associated non-thermal emission from the radio to $\gamma$ rays.
The multi-wavelength spectrum of Kes~73 is best described by a lepto-hadronic model with a total electron energy of $W_e = 2.4^{+1.5}_
{-0.8} \times 10^{47}$~erg, a magnetic field strength of $B = 25\,\mu$G, and a maximum electron energy of $1.1$~TeV. 
These parameters imply a synchrotron cutoff at $1.5 \times 
10^5$~GHz, well above our highest observed radio frequencies. 

This work sheds new light on the potential of high-resolution radio imaging to constrain the non-thermal CR spectrum from specific SNR regions. 
In particular, high-frequency single-dish radio data are crucial for assessing and identifying the spectral cutoff associated with the maximum energy of accelerated CR electrons. 
Significant progress in this type of study is expected thanks to the new 
multi-feed $Q$-band ($33-50$~GHz) and $W$-band receivers (up to $100$~GHz; \citealt{Navarrini_2016}) that are currently being implemented at the 
SRT. Radio observations at such high frequencies offer the opportunity to verify the cutoff frequency of the non-thermal spectrum associated with the western bright region of Kes~73, thereby enabling a definitive distinction between pure leptonic and hybrid lepto-hadronic models.

\begin{acknowledgements}

The Sardinia Radio Telescope is funded by the Ministry of University and Research (MUR), Italian Space Agency (ASI), and the Autonomous Region of Sardinia (RAS) and is operated as National Facility by the National Institute for Astrophysics (INAF). This publication makes use of data from FUGIN, FOREST Unbiased Galactic plane Imaging survey with the Nobeyama 45-m telescope, a legacy project in the Nobeyama 45-m radio telescope. We acknowledge the useful comments of the anonymous referee.
C.T. and G.U. acknowledge support from PRIN MUR 2022 (20224MNC5A), “Life, death and after-death of massive stars”, funded by European Union – Next Generation EU.

\end{acknowledgements}





%
\bibliographystyle{aa} 
\bibliography{AandA}
%
\label{LastPage}

\end{document}